\documentclass[fleqn,twoside]{article}
\usepackage{espcrc2}

\usepackage{amssymb}

\usepackage{graphicx}
\usepackage[figuresright]{rotating}

\newcommand{\AmS}{{\protect\the\textfont2
  A\kern-.1667em\lower.5ex\hbox{M}\kern-.125emS}}
  
\title{CMD-2 and SND results on the $\rho$, $\omega$ and $\phi$.}

\author{M.N. Achasov \address[BINP]{Budker Institute of Nuclear Physics,
                                    Siberian Branch of the Russian Academy of
                                    Sciences, \\ 11 Lavrentyev,Novosibirsk,
                                    630090, Russia}
\address[NSU]{Novosibirsk State University, 630090, Novosibirsk, Russia}
\thanks{E-mail:achasov@inp.nsk.su.}}

\begin{document}

\begin{abstract}
 The review of experimental results of the light vector mesons parameters
 studies with CMD-2 and SND detectors at VEPP-2M $e^+e^-$ collider in the
 energy region $360\leq\sqrt{s}\leq1380$ MeV is given.
 
\vspace{1pc}
\end{abstract}

\maketitle

\section{Introduction}

 During the last 35 years the cross section of the $e^+e^-\to hadrons$ 
 annihilation was studied in the energy region $\sqrt{s} \approx 0.4$ -- 190
 GeV at $e^+e^-$ colliders in CERN, DESY, KEK, SLAC, Cornell, Novosibirsk,
 Orsay, Frascati, Beijing. The energy range of the VEPP-2M collider
 (Novosibirsk, BINP SB RAS) \cite{vepp2} lays below 1.4 GeV in the beginning 
 of the energy scale. Besides VEPP-2M, the low energy region were studied in 
 experiments at ACO, DCI (Orsay) and ADONE (Frascati) $e^+e^-$ colliders. The
 VEPP-2M had finished its operation in year 2000 and now is being reconstructed
 into VEPP-2000 \cite{vepp2000}. At $B$-factory PEP-II (SLAC)  the cross 
 sections  of the processes $e^+e^-\to hadrons$ are measured using radiative
 return  method in the energy region from the reactions thresholds up to
 $\Upsilon(4S)$ mass \cite{babar-3pi,babar-4pi}. The $\phi$-factory DA$\Phi$NE
 (Frascati) produces experimental data in the narrow energy region around
 $\sqrt{s}\simeq 1020$ MeV. The radiative return method is also used there for
 the cross sections measurements below 1 GeV \cite{kloe-2pi}.
 
\begin{figure}
\includegraphics[scale=0.4]{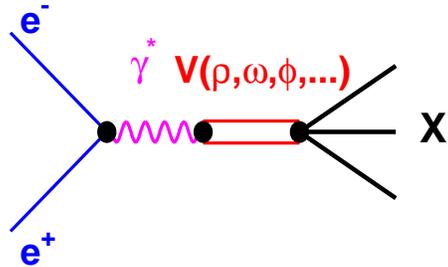}
\vspace{-2cm}
\caption{The $e^+e^- \to X$ transition diagram in VDM 
         (X denotes the final hadrons).}
\label{diag}
\end{figure}

 At the VEPP-2M energy region the $e^+e^-\to hadrons$ processes cross
 sections are described with 1\% accuracy by the vector mesons dominance model
 (VDM). In the VDM framework the virtual photon transits into the vector meson
 ($\rho,\omega,\phi$ or their excited states) with quantum numbers 
 $I^G(J^{PC})=1^+(1^{--})$ or  $0^-(1^{--})$, which in its turn decays to
 hadrons (fig.\ref{diag}).

 Studies of the $e^+e^-\to hadrons$ processes allow to determine parameters of 
 the vector mesons $\rho,\omega,\phi$ and their excitations, provide
 information about interference between mesons and reactions dynamics.

\section{Experiment}

 From 1995 to 2000 the CMD-2 \cite{kmd2} and SND \cite{sndnim} detectors 
 simultaneously operated at the VEPP-2M $e^+e^-$ collider in the energy range
 $\sqrt{s}$ from 360 to 1400 MeV.  The VEPP-2M luminosity varies in the range 
 from $\sim 10^{28}$cm$^{-2}$s$^{-1}$ at $\sqrt{s} = 360$ MeV to 
 $\sim 10^{30}$cm$^{-2}$s$^{-1}$ at $\sqrt{s}=1$ GeV. The total 
 integrated luminosity collected by each detector is about 30 pb$^{-1}$.

 SND detector contains several subsystems. The tracking system includes two
 cylindrical drift chambers. The three-layer electromagnetic calorimeter is
 based on NaI(Tl) crystals. The muon/veto system consists of plastic 
 scintillation counters and two layers of streamer tubes. 
 CMD-2 detector contains tracking system which includes a cylindrical drift
 chamber and double-layer multiwire proportional chamber ($z$-chamber). The
 photons energy and angles are measured by using the barrel CsI and endcup BGO
 calorimeters installed inside the superconductive solenoid with a 1 T
 magnetic field. The muon system consists of two layers of streamer tubes.
 The detailed descriptions of the detectors are given in 
 Ref.\cite{kmd2,sndnim}.

 CMD-2 and SND can detect and completely reconstruct events containing both
 neutral and charged particles.

 The experiments were performed using energy scan method. After events
 reconstruction and analysis the cross section of the process 
 $e^+e^-\to hadrons$ is obtained as 
\begin{equation}
\sigma = {N \over IL \mbox{~} \varepsilon
                     \mbox{~} \delta_{rad}
                     \mbox{~} \delta_{s} },
\end{equation}
 where $N$ is the process events number, $IL$ is the integrated luminosity 
 (measured by using reactions with well known cross sections:
 $e^+e^-\to e^+e^-$ and $\gamma\gamma$), $\varepsilon$ is the detection 
 efficiency (obtained from Monte-Carlo simulation), $\delta_{rad}$ is the 
 radiative correction which takes into account the photons emission by initial
 and in some cases by final state particles, $\delta_{s}$ is the correction
 due to the beam energy spread.
 
 In order to obtain the vector mesons parameters the measured cross section
 of the process $e^+e^-\to X$, where $X$ denotes the final hadron system, is
 fitted by theoretical expression:

\begin{eqnarray}
 \sigma(s) & = & {4\pi\alpha \over s^{3/2}} P(s) \biggl|
 \sum_{V=\rho,\omega,\phi, {\ldots} }
 { \Gamma_Vm_V^3\sqrt{m_V\sigma(V\to X)} \over D_V(s)} \nonumber \\
 & & { e^{i\phi_V}\over\sqrt{P(m_V)}}
 \biggr|^2.
\end{eqnarray}
 Here $P_X(s)$ is a phase space factor including dynamics of the vector meson
 transition to the final state, $m_V$ and $\Gamma_V$ are the meson mass and
 full width, $\sigma(V\to X)=12\pi B(V\to X)B(V\to e^+e^-)/m_V^2$,
 $D_V(s)$ is the vector meson $V$ propagator, $\phi_V$ is the interference
 phase. The parameters $m_V$, $\Gamma_V$, $B(V\to X)B(V\to e^+e^-)$, $\phi_V$
 are obtained form the fit.

 CMD-2 and SND experiments have obtained quite important  results on 
 the $\rho,\omega,\phi$ rare decays (relative probabilities less than
 $10^{-3}$). The well known decays $\rho\to\pi^0\gamma$ 
 \cite{snd-pi0g,kmd-pi0g},  $\rho\to\eta\gamma$ É $\omega\to\eta\gamma$  
 \cite{kmd-pi0g,kmd-etag,snd-etag,snd-hep05}, $\omega\to\pi^0\pi^0\gamma$ 
 \cite{snd-roppg1,snd-roppg2,kmd-roppg}, $\phi\to\pi^+\pi^-$ \cite{phi2pi}
 were also studied. The eight decays
 $\phi\to f_0\gamma$ \cite{f0g1,f0a0,f0g2,f0a02,f0g3}, 
 $\phi\to a_0\gamma$ \cite{a0g1,f0a0,a0g2,f0a02}, 
 $\phi\to\eta^\prime\gamma$ \cite{phietap1,phietap2,phietap3,phietap4,phietap5},
 $\phi\to\omega\pi^0$ \cite{phiomp1,phiomp2,phiomp3}, 
 $\phi\to\pi^+\pi^-\pi^+\pi^-$ \cite{phi4pic}, 
 $\rho\to\pi^0\pi^0\gamma$ \cite{snd-roppg1,snd-roppg2,kmd-roppg}, 
 $\rho\to\pi^+\pi^-\pi^+\pi^-$ \cite{rho4pic}, 
 $\rho\to\pi^+\pi^-\pi^0$ \cite{snd-3pi4}, $\phi\to\pi^0 e^+e^-$
 \cite{kmd-pioee} were observed for the first time. In this paper the vector
 mesons main decays and parameters studies are  reviewed.
 
\section{Cross sections of $e^+e^-\to hadrons$ annihilation}

\begin{figure}[t]
\includegraphics[scale=0.5]{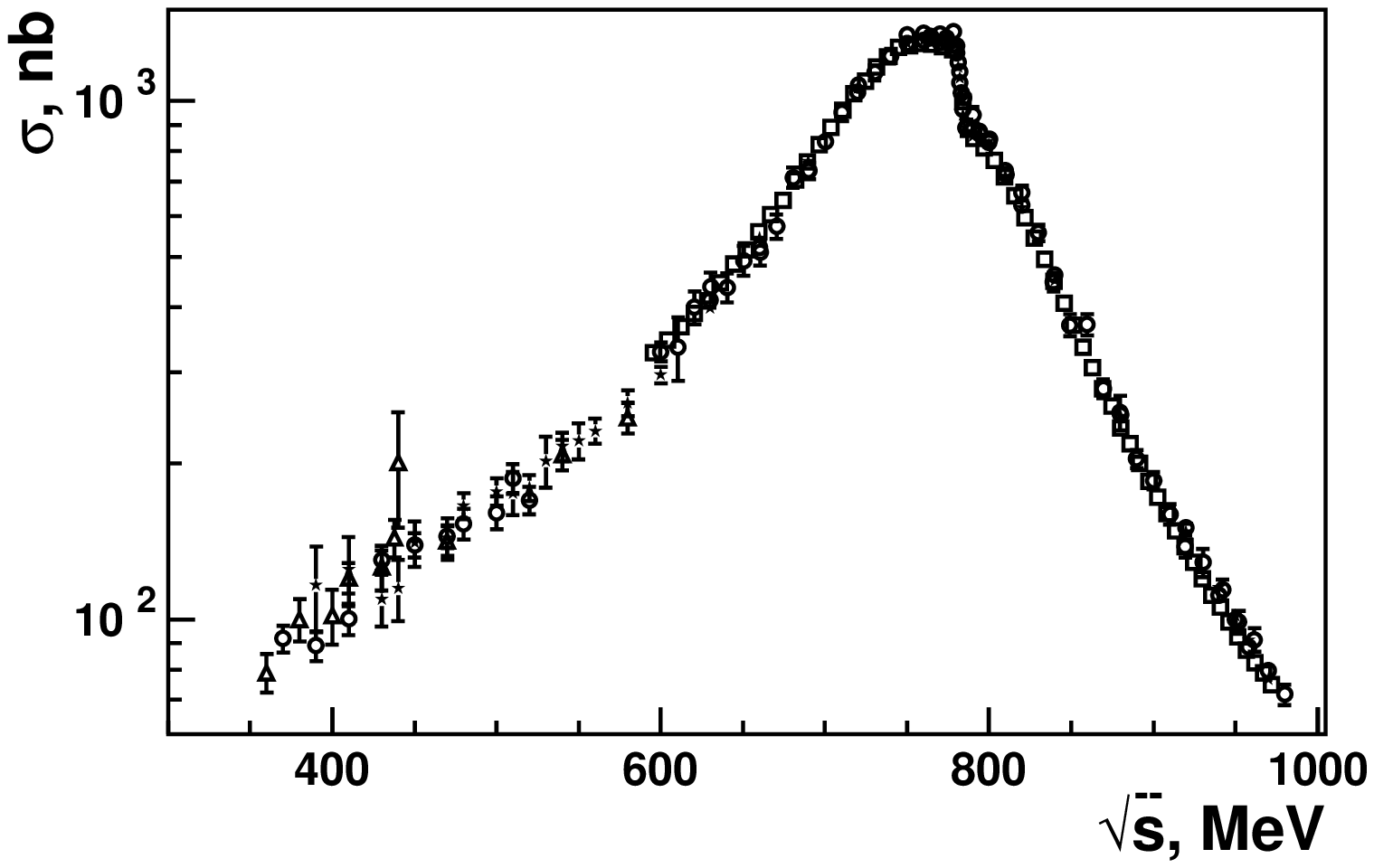}
\vspace{-1.5cm}
\caption{The $e^+e^-\to\pi^+\pi^-$ cross section. OLYA and CMD
         ($\vartriangle$) \cite{olya}, SND ($\star$) \cite{snd-2pi}, CMD-2 
	 ($\circ$)  \cite{tau04,hep05,hadr05} and KLOE ($\square$) 
	 \cite{kloe-2pi} data are shown.}
\label{2pi-1}
\end{figure}

\begin{figure}[t]
\includegraphics[scale=0.5]{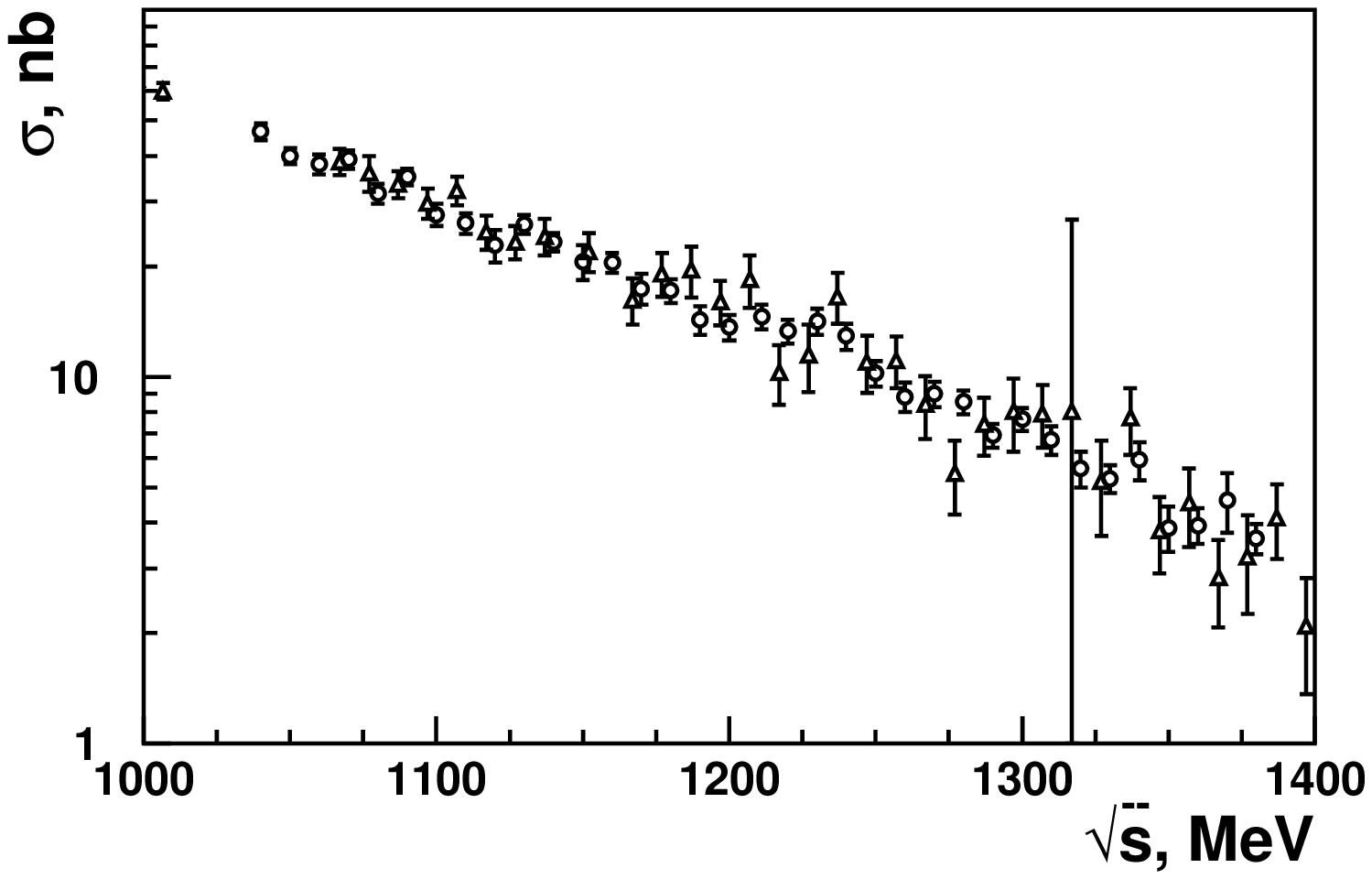}
\vspace{-1.5cm}
\caption{The $e^+e^-\to\pi^+\pi^-$ cross section. OLYA ($\vartriangle$) 
         \cite{olya} and CMD-2 ($\circ$) \cite{kmd-2pi2} data are shown.}
\label{2pi-2}
\end{figure}

 The $e^+e^-\to\pi^+\pi^-$ cross section in the VEPP-2M energy region
 measured in various experiments 
 \cite{olya,snd-2pi,tau04,hep05,hadr05,kmd-2pi1,kmdrean,kloe-2pi,kmd-2pi2} is
 shown in fig.\ref{2pi-1}, \ref{2pi-2}. For the data description the 
 $\rho,\omega,\rho^\prime$ mesons are required. The CMD-2 data have 
 0.6 -- 0.8 \% systematic error below and 1.2 -- 4.2 \% above 1 GeV. SND 
 measured the cross section below 1 GeV with accuracy 1.3 -- 3.4  \%. SND and
 CMD-2 results are in agreement with each other and with the CMD and OLYA 
 measurements. However they are in conflict with KLOE measurement.
 
\begin{figure}
\includegraphics[scale=0.6]{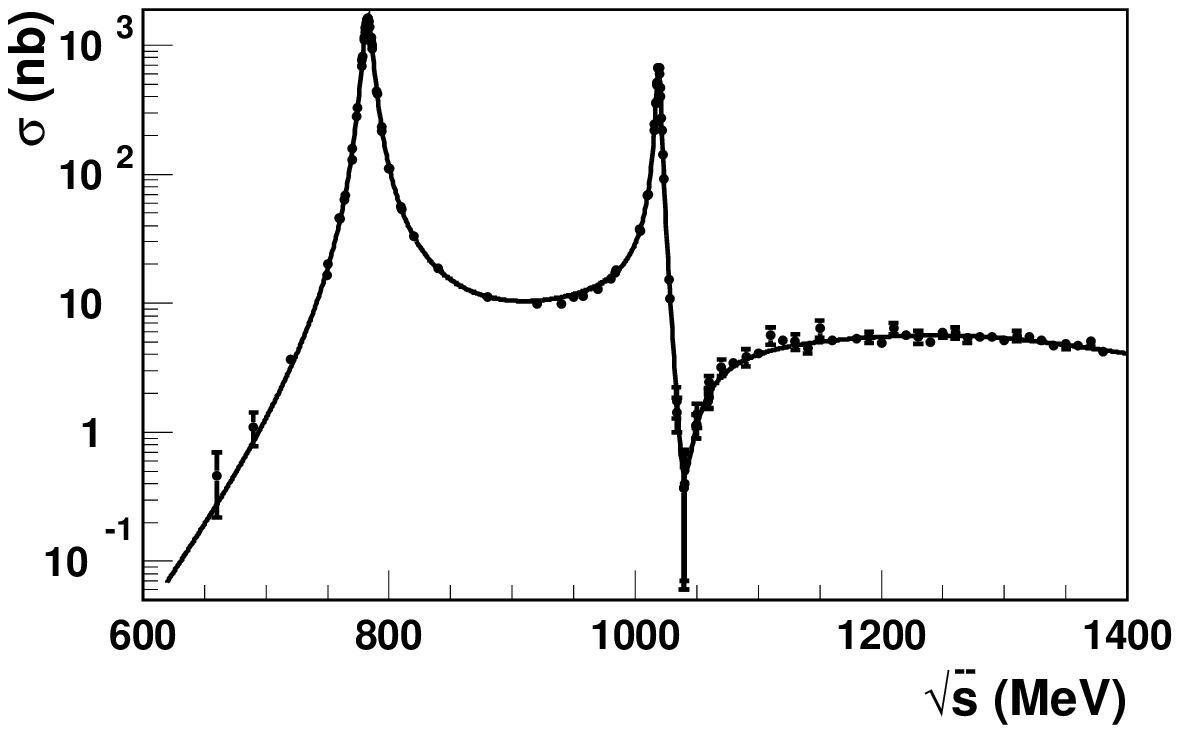}
\vspace{-1.5cm}
\caption{The $e^+e^-\to\pi^+\pi^-\pi^0$ cross section measured by SND
         \cite{snd-3pi1,snd-3pi3,snd-3pi4}. The curve is the fit with the
	 $\omega,\phi,\rho,\omega^\prime,\omega^{\prime\prime}$ resonances.}
\label{pi3-1}
\includegraphics[scale=0.6]{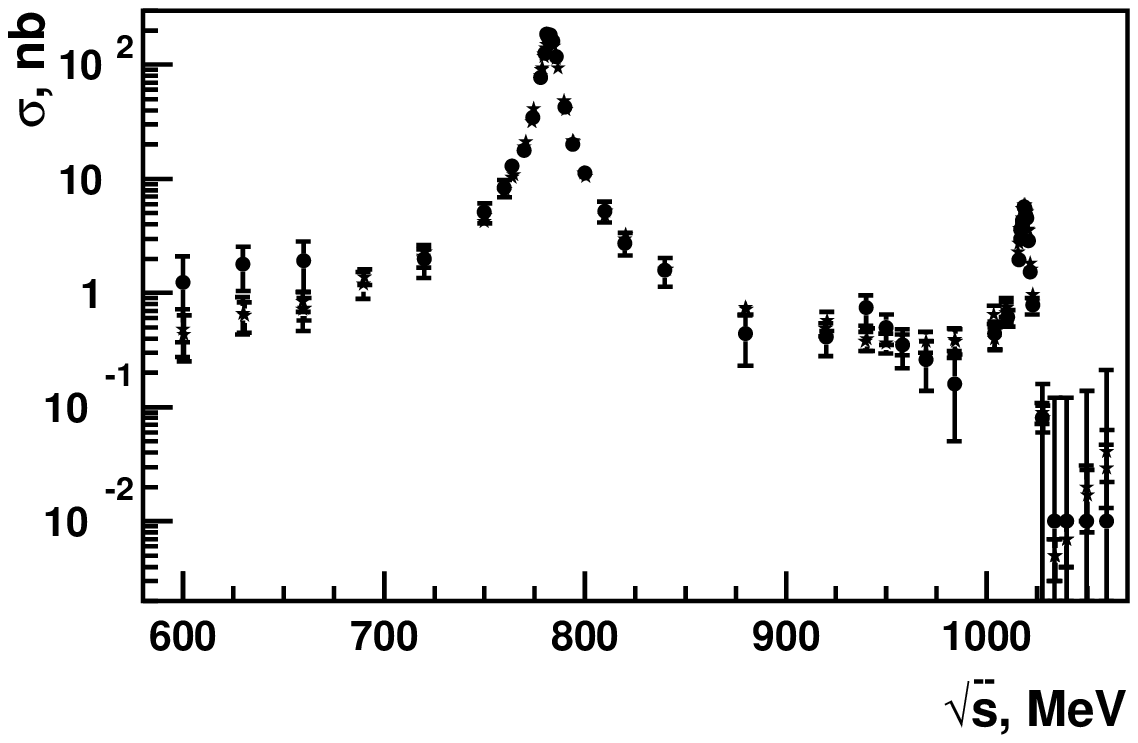}
\vspace{-1.5cm}
\caption{The $e^+e^-\to\pi^0\gamma$ cross section. The SND
         ($\star$) \cite{snd-pi0g,snd-hep05} and CMD-2 ($\bullet$) data are
         shown \cite{kmd-pi0g}.} 
\label{pi0g}
\includegraphics[scale=0.6]{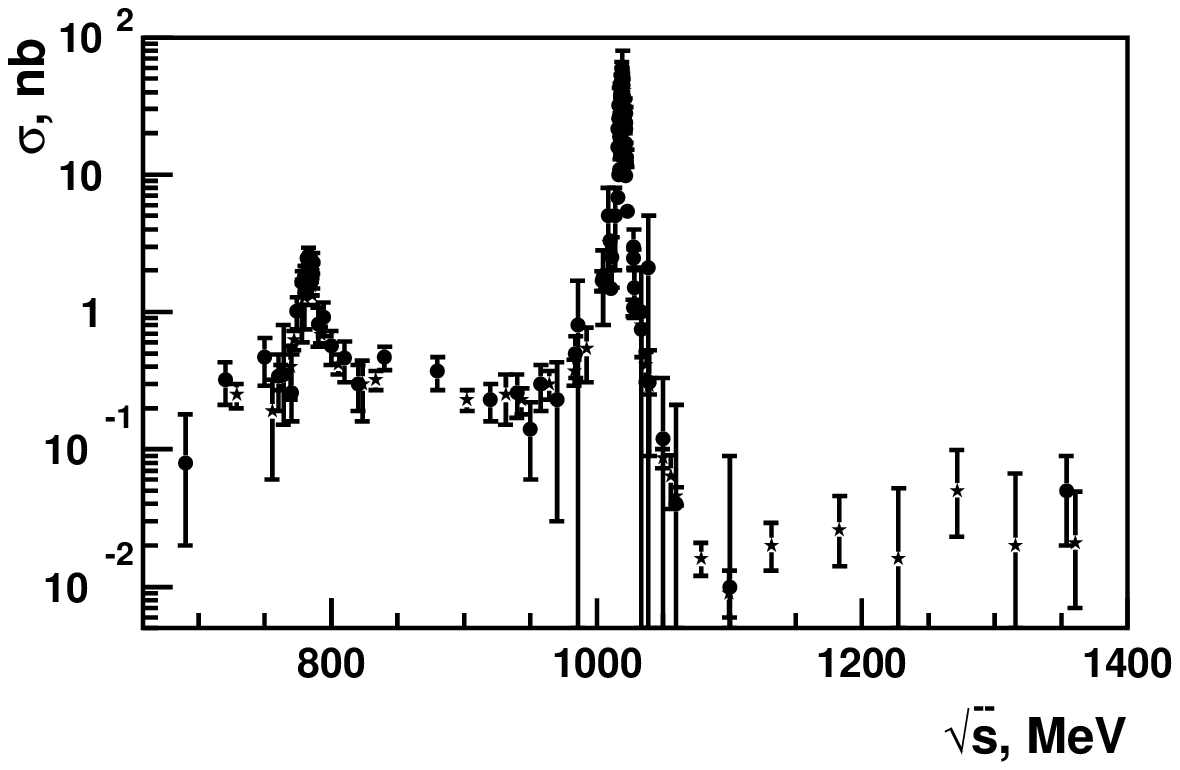}
\vspace{-1.5cm}
\caption{The $e^+e^-\to\eta\gamma$ cross section. The SND
         ($\star$) \cite{snd-hep05} and CMD-2 ($\bullet$) data are shown
	 \cite{kmd-etag}.} 
\label{etag}
\end{figure}

 The $e^+e^-\to\pi^+\pi^-\pi^0$ process was studied with SND in the energy
 region $\sqrt{s}$ from 600 to 1380 MeV \cite{snd-3pi4,snd-3pi1,snd-3pi2,snd-3pi3} 
 (fig.\ref{pi3-1}). The curve is the result of the fit taking into
 account $\omega,\phi,\rho,\omega^\prime$ and $\omega^{\prime\prime}$. The
 systematic accuracy of the measurement is 3.4 -- 5.4 \%. In experiments at
 VEPP-2M the clear evidence of the broad resonance structure at $\sqrt{s}=1100$
 -- 1400 MeV, which is identified as $\omega^\prime$ resonance, was obtained.
 SND measurements were confirmed by BABAR radiative return method result
 \cite{babar-3pi}. The $e^+e^-\to\rho\pi\to\pi^+\pi^-\pi^0$ transition
 dominates in this reaction \cite{snd-3pi2,snd-3pi3,kmd-3pi2,kloe-3pi}. In the
 energy region above the $\phi$-meson mass the transition
 $e^+e^-\to\omega\pi^0\to\pi^+\pi^-\pi^0$, predicted in Ref. \cite{tomp3pi}, 
 was observed and studied \cite{snd-3pi3}. The intermediate
 state different from these two (maybe 
 $e^+e^-\to\rho^\prime\pi\to\pi^+\pi^-\pi^0$) was observed by KLOE and CMD-2
 in the $\phi$-meson energy region \cite{kloe-3pi,pi3epif}. CMD-2 has reported
 the cross section measurements in the vicinity of the $\phi$ and $\omega$
 peaks \cite{kmd-phi,kmd-3pi2,kmd-3pi3,kmdrean} based on a part of available
 statistic. The data analysis is going on \cite{pi3epif,pi3gorb}.
\begin{figure}
\includegraphics[scale=0.6]{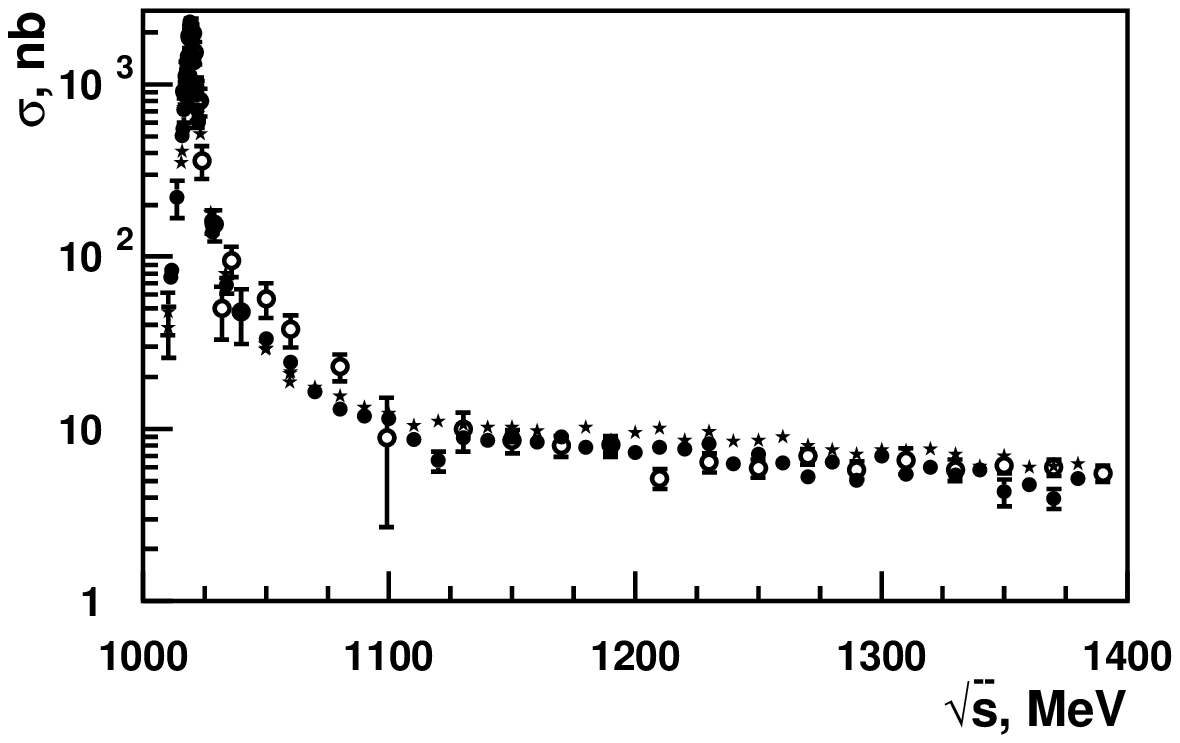}
\vspace{-1.5cm}
\caption{The $e^+e^-\to K^+K^-$ cross section. The OLYA ($\circ$)
         \cite{olya-kkc}, SND ($\star$) \cite{snd-3pi1,snd-hep05} and CMD-2 
	 ($\bullet$) \cite{kmd-phi,kmd-kkc} data are shown.}
\label{kkc-1}
\includegraphics[scale=0.6]{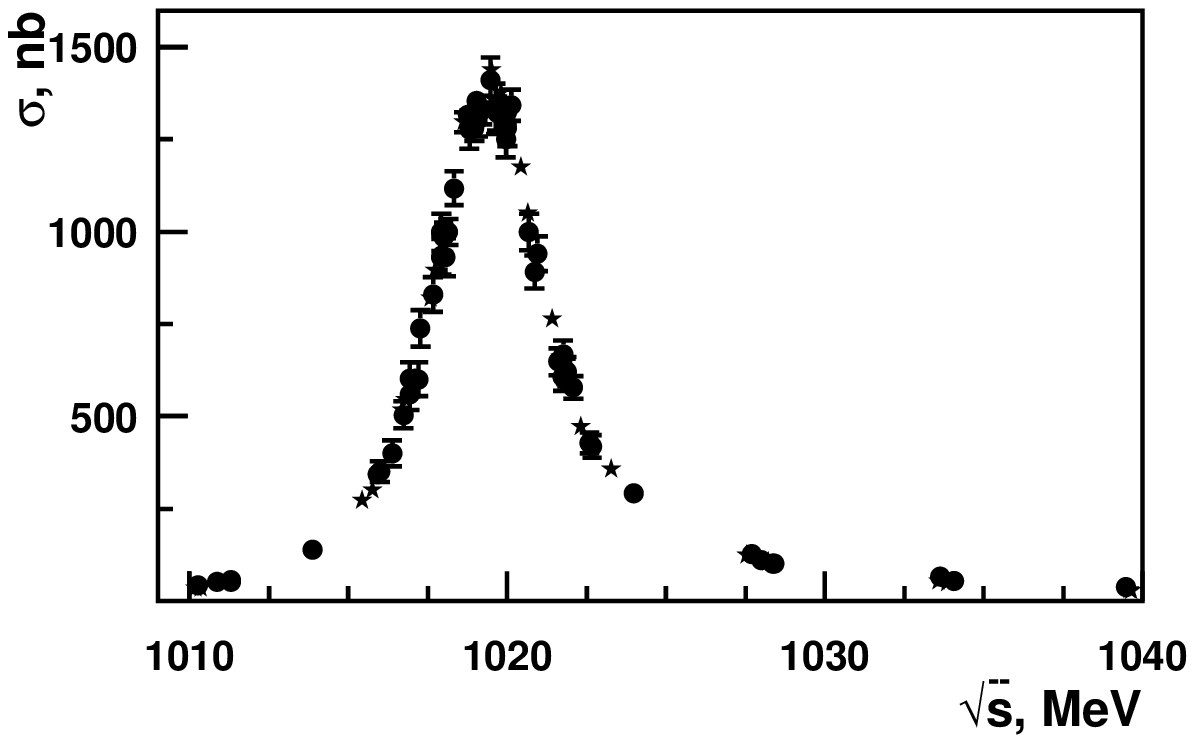}
\vspace{-1.5cm}
\caption{The $e^+e^-\to K_SK_L$ cross section. The SND ($\star$)
         \cite{snd-3pi1} and CMD-2 ($\bullet$) \cite{kmd-ksl} are shown.}
\label{ksl-1}
\includegraphics[scale=0.6]{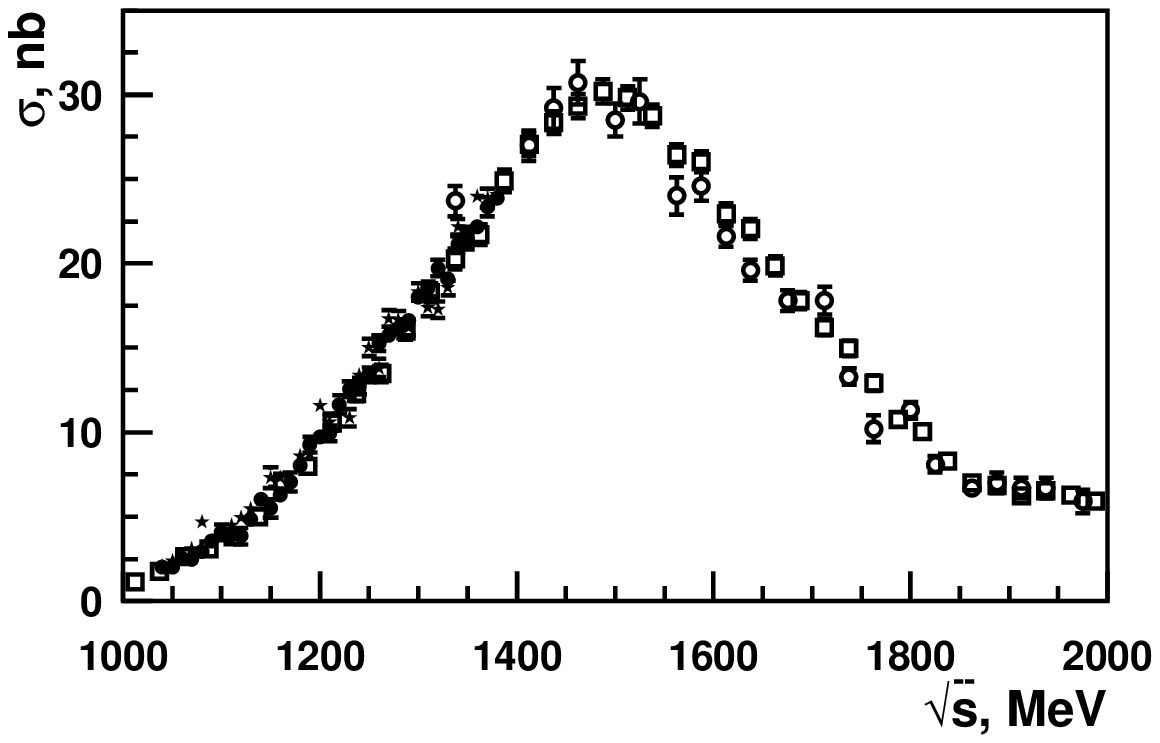}
\vspace{-1.5cm}
\caption{The $e^+e^-\to\pi^+\pi^-\pi^+\pi^-$ cross section. The 
         DM2 ($\circ$) \cite{dm2-4pi}, SND ($\star$) \cite{snd-4pi}, 
         CMD-2 ($\bullet$)  \cite{kmd-4pi} and BABAR ($\square$) 
	 \cite{babar-4pi} data are shown.}
\label{4pic}
\end{figure}

 CMD-2 and SND measured the cross sections of the  $e^+e^-\to\pi^0\gamma$ and
 $e^+e^-\to\eta\gamma$ processes. The systematic errors of the CMD-2 and SND
 measurements of the $e^+e^-\to\pi^0\gamma$ cross section are 6 \% and 3 \%
 respectively \cite{kmd-pi0g,snd-pi0g,snd-hep05,snd-pi0g2}. The energy
 dependence of the cross section (fig.\ref{pi0g}) is described by $\rho,\omega,\phi$ mesons
 contributions. The  $e^+e^-\to\eta\gamma$ process was studied by using events
 with the $\eta\to\pi^0\pi^0\pi^0,\pi^+\pi^-\pi^0,\gamma\gamma$ decays
 \cite{kmd-pi0g,kmd-etag,snd-etag,snd-hep05,snd-pi0g2,snd-etag3pc,snd-etag7g,kmd-phi,kmd-etag3pc}.
 The systematic errors depend on energy and at the $\phi$-meson region they
 are about 3-7 \%. The cross section energy dependence is shown in 
 fig.\ref{etag}. For the data description the $\phi,\rho,\omega,\rho^\prime$
 resonances are needed. 

 The $e^+e^-\to K^+K^-$ process cross section measured by SND 
 \cite{snd-3pi1,snd-hep05}, CMD-2 \cite{kmd-phi,kmd-kkc} and OLYA
 \cite{olya-kkc} in the energy region $\sqrt{s}$ below 1380 MeV is shown in
 fig.\ref{kkc-1}. In the energy region $\sqrt{s}>1060$ MeV the SND and CMD-2
 data are preliminary. The systematic accuracies of the CMD-2 and SND 
 measurements in the $\phi$-meson energy range are 3-4 and 7 \% respectively.

 The $e^+e^-\to K_SK_L$ cross section was measured by using $K_S\to\pi^0\pi^0$
 and $\pi^+\pi^-$ decays. The cross section measured by SND \cite{snd-3pi1}
 and CMD-2 \cite{kmd-ksl,kmdrean} in the vicinity of the $\phi$-meson 
 (fig.\ref{ksl-1}) has systematic errors 3.3 \%  and 1.7 \% respectively. The
 results of different experiments are in agreement. The measurements above 
 $\phi$-meson were reported by SND and CMD-2 in 
 Ref.\cite{snd-hep05,kmd-ksl2,kmd-ksl3}.
 
 The $e^+e^-\to K\overline{K}$ data near the $\phi(1020)$-meson peak can be
 described by taking into account the $\phi,\rho,\omega$ amplitudes only. The 
 $\rho$ and $\omega$ coupling constants agree with the naive quark model
 predictions. At higher energies the $\phi^\prime$ contribution should be 
 added to the reactions amplitudes. 

 The $e^+e^-\to\pi^+\pi^-\pi^+\pi^-$ and $e^+e^-\to\pi^+\pi^-\pi^0\pi^0$ 
 processes have the largest cross sections in the energy region 
 $\sqrt{s}>1$ GeV . The CMD-2 \cite{kmd4p}  and CLEO-2 \cite{cleo4p} had shown 
 that in the VEPP-2M energy region in the $e^+e^-\to\pi^+\pi^-\pi^+\pi^-$ 
 process the $\rho\pi\pi$, and in the $e^+e^-\to\pi^+\pi^-\pi^0\pi^0$ the 
 $\rho\pi\pi$ and $\omega\pi^0$ mechanisms are dominant. Moreover the dynamics
 of the  $e^+e^-\to\rho\pi\pi$ reaction can be described with $a_1\pi$
 intermediate state. The SND analysis confirmed these conclusions 
 \cite{snd-4pi}. The $e^+e^-\to\pi^+\pi^-\pi^+\pi^-$ cross section measured by
 DM2 \cite{dm2-4pi},  SND \cite{snd-4pi}, CMD-2 \cite{kmd-4pi} and BABAR 
 \cite{babar-4pi} is shown in fig.\ref{4pic}. The results agree with each
 other. The $e^+e^-\to\omega\pi^0$ was steadied in two final states
 $e^+e^-\to\omega\pi^0\to\pi^+\pi^-\pi^0\pi^0$ and
 $e^+e^-\to\omega\pi^0\to\pi^0\pi^0\gamma$ by both SND \cite{snd-4pi,snd-ppg}
 and CMD-2 \cite{hadr05,kmd-ppg}. These results are in agreement also.
 
 The fit to the experimental data shows that within the measurement accuracy
 the $e^+e^-\to hadrons$ cross sections can be described by VDM taking into 
 account $\rho,\omega,\phi$-mesons and their excited states
 $\rho^\prime,\omega^\prime,\phi^\prime,{\ldots}$ .

\section{Light vector mesons parameters}

\begin{sidewaystable}
\caption{The average values of the $\rho,\omega,\phi$ mesons parameters
         measured by CMD-2 and SND in comparison with other experimental 
	 results.} 
\label{tab1}
\begin{tabular}{lll}
\hline
 & CMD-2 and SND                               & other data \\ \hline
$m_\rho$, MeV      & $775.0\pm 0.5$ \cite{kmd-2pi1,kmdrean,snd-2pi,snd-3pi2} & 
                     $775.7\pm 0.4$ \cite{cleo2,aleph2,olya,kloe-3pi} \\
$\Gamma_\rho$, MeV & $145.5\pm 1.1$ \cite{kmd-2pi1,kmdrean,snd-2pi,snd-3pi2} &
                     $148.9\pm 0.8$ \cite{cleo2,aleph2,olya,kloe-3pi} \\
$\Gamma(\rho\to e^+e^-)$, keV & $7.09\pm 0.08$ \cite{kmd-2pi1,kmdrean,snd-2pi} & 
                                $6.65\pm 0.29$ \cite{olya,ospk72} \\
\hline
$m_\omega$, MeV      & $782.78\pm 0.07$ \cite{kmd-3pi3,kmdrean,snd-3pi4} & 
                       $782.30\pm 0.12$ 
		   \cite{spec95,cbar1,cbar2,aste,olyaomega,cntr76,dm1} \\
$\Gamma_\omega$, MeV & $8.68\pm 0.13$ \cite{kmd-3pi3,kmdrean,snd-3pi4} & 
$8.41\pm 0.09$ \cite{spec95,nd,cmdgom,olyaomega,dm1,ospk72}  \\
$B(\omega\to\pi^+\pi^-\pi^0)B(\omega\to e^+e^-)\times 10^{5}$&
$6.37\pm 0.12$ \cite{kmd-3pi3,kmdrean,snd-3pi4} &
$6.42\pm 0.14$ \cite{babar-3pi,nd,cmdgom,olyaomega,dm1,ospk72} \\
$B(\omega\to\pi^0\gamma)B(\omega\to e^+e^-)\times 10^{6}$&
$6.49\pm 0.2 \cite{snd-pi0g,kmd-pi0g}$&$6.34\pm 0.3$ \cite{nd} \\
$B(\omega\to\pi^+\pi^-)$ &$0.0156\pm 0.001$ \cite{kmd-2pi1,kmdrean,snd-2pi}&
$0.0227\pm 0.003$ \cite{olya,dm1-2,ospk72,aspk72,aspk71}\\
\hline
$m_\phi$, MeV      & $1019.45\pm 0.02$ \cite{kmd-kkc,snd-3pi1,kmd-ksl}
& $1019.47\pm 0.07$ \cite{omeg98,mpsf86,arg85,aems82,olya78} \\
$\Gamma_\phi$, MeV & $4.25\pm 0.02$ \cite{kmd-kkc,kmd-ksl,snd-3pi1} & 
$4.28\pm 0.17$ \cite{aems82,olya-kkc,dm1,olya78,cntr74,ospk71,ospk70} \\
$B(\phi\to K^+K^-)B(\phi\to e^+e^-)\times 10^{5}$ &
$14.34\pm 0.34$ \cite{kmd-kkc,snd-3pi1,kmd-phi}&
$15.33\pm 0.54$ \cite{ospk71,ospk70,hbc76,hbc74,hbc66}\\
$B(\phi\to K_SK_L)B(\phi\to e^+e^-)\times 10^{5}$&
$10.00\pm 0.14$ \cite{kmdrean,kmd-ksl,snd-3pi1,kmd-phi}&
$10.13\pm 0.30$ \cite{nd,olya78,ospk74,ospk71,ospk70,hbc78,hbc771,hbc772,hbc72} \\
$B(\phi\to\pi^+\pi^-\pi^0)B(\phi\to e^+e^-)\times 10^{5}$&
$4.56\pm 0.16$ \cite{snd-3pi1,kmd-3pi2,kmd-phi}&
$4.51\pm 0.11$ \cite{babar-3pi,nd,dm1,olya78,ospk76,ospk74}\\
$B(\phi\to\eta\gamma)B(\omega\to e^+e^-)\times 10^{6}$&
$3.88\pm 0.06$ \cite{kmd-pi0g,kmd-etag,snd-hep05,snd-pi0g2,snd-etag7g,kmd-etag3pc}
&$3.96\pm 0.18$ \cite{nd,olya83,ospk74,cntr77}\\
$B(\phi\to e^+e^-)\times 10^{4}$ &
$2.90\pm 0.04$ \cite{kmd-kkc,snd-3pi1,snd-2mu,kmd-phi}&
$3.10\pm 0.05$ \cite{kloe-ll}\\
\hline
\end{tabular}
\end{sidewaystable}

\begin{table*}
\caption{The $\omega^\prime$ and $\omega^{\prime\prime}$ parameters obtained
         from the SND and DM-2 data analysis \cite{snd-3pi3,snd-3pi4}.}
\label{tab2}
\begin{tabular}{@{}llllll}
\hline
 V &$m_V$, MeV& $\Gamma_V$, MeV& $\sigma(V\to 3\pi$), nb&
 $\sigma(V\to \omega\pi\pi$, nb& $\Gamma(V\to e^+e^-)$, Ü÷ \\
\hline
$\omega^\prime$&$1400 \pm 50 \pm 130$&$870 \pm^{500}_{300} \pm 450$&
$4.9 \pm 1.0 \pm 1.6$ & &$\sim 570$ \\
$\omega^{\prime\prime}$&$1770\pm 50\pm 60$ & $490\pm^{200}_{150}\pm 130$&
$5.4 \pm^{2.0}_{0.4} \pm 3.9$ & $1.9 \pm 0.4 \pm 0.6$ &$\sim 860$ \\
\hline
\end{tabular}
\end{table*}

 The parameters of the light vector mesons $\rho,\omega,\phi$ (Tab..\ref{tab1})
 were extracted from the measured cross sections. The CMD-2 and SND averaged 
 values together with the results of the other measurements are shown in 
 Tab.\ref{tab1}. The $\rho$-meson parameters were obtained from the study of 
 the $e^+e^-\to\pi^+\pi^-$ cross section. The accuracy of the
 $\rho\to e^+e^-$ decay width in the recent VEPP-2M experiments was improved
 by 3 times in comparison with previous measurements \cite{olya,ospk72}.

 The $\omega$-meson main parameters were obtained from the
 $e^+e^-\to\pi^+\pi^-\pi^0,\pi^0\gamma$ and $\pi^+\pi^-$ processes cross
 sections analysis. The results of the CMD-2 and SND have accuracy compatible
 or better than the other experiments. In particular, the accuracy of the
 $G$-parity suppressed decay $\omega\to\pi^+\pi^-$ was improved by factor 3.

 The accuracy of the parameters  $B(\phi\to e^+e^-)B(\phi\to\eta\gamma)$ and
 $B(\phi\to e^+e^-)B(\phi\to K_SK_L)$ was improved also. Using the CMD-2 and 
 SND results \cite{snd-3pi1,snd-2mu,kmd-kkc,kmd-ksl,kmdrean,pi3epif,kmd-etag}
 the $B(\phi\to e^+e^-)=(2.90\pm 0.04)\times 10^{-4}$ was obtained. This value
 deviates from the KLOE result 
 $B(\phi\to e^+e^-)=(3.10\pm 0.05)\times 10^{-4}$  \cite{kloe-ll} by three
 standard deviations.

 The light vector mesons $\rho,\omega,\phi$ are studied rather well. About
 $\omega^\prime, \rho^\prime, \phi^\prime, {\ldots} $ we know with certainty
 that such resonances exist. However their parameters are not well established
 and their nature is not clear due to poor accuracy of experimental data,
 large width of these states and model uncertainty in their description. The parameters of the
 $\omega^\prime$ and $\omega^{\prime\prime}$ were estimated from combined
 analysis of the $e^+e^-\to\pi^+\pi^-\pi^0$ process cross section measured by
 SND and of the DM-2 data on the  $e^+e^-\to\omega\pi\pi$ reaction
 (Tab.\ref{tab2}). Using the leptonic widths obtained from the fit the 
 following ratios can be obtained in the framework of the
 nonrelativistic quark model:
\begin{eqnarray}
 \biggl| {{\Psi_{\omega^{\prime}}^S(0)}\over{\Psi_\omega^S(0)}}\biggr|^2 =
 \biggl({{m_{\omega^\prime}}\over{m_\omega}} \biggr)^2 \cdot
 {{\Gamma(\omega^\prime\to e^+e^-)}\over{\Gamma(\omega\to e^+e^-)}} \sim 4,
\end{eqnarray}
\begin{eqnarray}
 \biggl|
 {{\Psi_{\omega^{\prime\prime}}^S(0)}\over{\Psi_\omega^S(0)}}\biggr|^2
 = \biggl({{m_{\omega^{\prime\prime}}}\over{m_\omega}} \biggr)^2 \cdot
 {{\Gamma(\omega^{\prime\prime}\to e^+e^-)}\over{\Gamma(\omega\to e^+e^-)}}
 \sim 5,
\end{eqnarray}
 where $\Psi_V^S(0)$ is the radial wave function of the $q\overline{q}$ bound
 state at the origin. These ratios are about 10 times higher than analogous
 values for the $c\overline{c}$ and $b\overline{b}$ states:
 $|\Psi_{\psi(2S)}^S(0)/\Psi_{J/\psi}^S(0)|^2\simeq 0.57$,
 $|\Psi_{\Upsilon(2S)}^S(0)/\Psi_{\Upsilon(1S)}^S(0)|^2\simeq 0.44$,
 $|\Psi_{\Upsilon(3S)}^S(0)/\Psi_{\Upsilon(1S)}^S(0)|^2\simeq 0.43$.
 These results can indicate to the unusual nature of the
 $\omega^\prime,\omega^{\prime\prime}$. More precise data and deeper
 analysis are required however, to draw final conclusions.

\section{Conclusion}

 In the 1995-2000 the experiments with CMD-2 and SND detectors at VEPP-2M were
 fulfilled. The cross section of the $e^+e^-$ annihilation in hadrons were
 measured in the energy region $\sqrt{s}=360-1380$ MeV. Results of these
 experiments determine nowadays the accuracy of the light vector mesons
 parameters determination. They are one of the main source of information
 about particle physics at low energies.

 The author is grateful to V.P. Druzhinin, S.I. Serednyakov and Z.K. Silagadze
 for useful discussions. The work is supported in part by grants 
 Sci.School-1335.2003.2, RFBR 04-02-16181-Á, 04-02-16184-Á, 05-02-16250-Á,
 06-02-16192-a.


\begin{thebibliography}{99}
\bibitem{vepp2}
 A.N. Skrinsky, in Proceedings of Workshop on physics and detectors for
 DA$\Phi$NE, Frascati, Italy, April 4-7, 1995, p.3.
\bibitem{vepp2000}
 Yu.M.Shatunov et al, Project of a new electron-positron collider VEPP-2000,
 in Proc. of the 2000 European Particle Acc. Conf., Vienna (2000), p.439
\bibitem{babar-3pi}
 B. Aubert et al., Phys. Rev. D 70 (2004) 072004.
\bibitem{babar-4pi}
 B. Aubert et al., Phys. Rev. D 71 (2005) 052001.
\bibitem{kloe-2pi}
 A. Aloisio et al., Phys. Lett. B 606 (2005) 12.
\bibitem{kmd2}
 R.R. Akhmetshin et al., Preprint 99-11, Budker INP ( Novosibirsk, 1999).
\bibitem{sndnim}
 M.N. Achasov et al., Nucl. Instr. and Meth. A 449 (2000) 125.
\bibitem{snd-pi0g}
 M.N. Achasov et al., Phys. Lett. B 559 (2003) 171.
\bibitem{kmd-pi0g}
 R.R. Akhmetshin et al., Phys. Lett. B 605 (2005) 26.
\bibitem{kmd-etag}
 R.R. Akhmetshin et al., Phys. Lett. B 509 (2001) 217.
\bibitem{snd-etag}
 M.N. Achasov et al., Pisma Zh. Eksp. Teor. Fiz. 72 (2000) 411.
\bibitem{snd-hep05}
 S.I. Serednyakov et al., in Proceedings of International European Conference
 on High Energy Physics, Lisboa, Portugal, July 21-27, 2005; hep-ex/0512027.
\bibitem{snd-roppg1} 
 M.N. Achasov et al., Pisma Zh. Eksp. Teor. Fiz. 71 (2000) 519.
\bibitem{snd-roppg2}
 M.N. Achasov et al., Phys. Lett. B 537 (2002) 201.
\bibitem{kmd-roppg}
 R.R. Akhmetshin et al., Phys. Lett. B 580 (2004) 119.
\bibitem{phi2pi}
 M.N. Achasov et al., Phys. Lett. B 474, 188 (2000).
\bibitem{f0g1}
 V.M. Aulchenko et al., Phys. Lett. B 440 (1998) 442.
\bibitem{f0a0}
 M.N. Achasov et al., Yad. Fiz. 62, 486 (1999).
\bibitem{f0g2}
 M.N. Achasov et al., Phys. Lett. B 485 (2000) 349.
\bibitem{a0g1}
 M.N. Achasov et al., Phys. Lett. B 438 (1998) 441.
\bibitem{a0g2}
 M.N. Achasov et al., Phys. Lett. B 479 (2000) 53.
\bibitem{f0a02}
 R.R. Akhmetshin et al., Phys. Lett. B 462 (1999) 380.
\bibitem{f0g3}
 R.R. Akhmetshin et al., Phys. Lett. B 462  (1999) 371.
\bibitem{phietap1}
 R.R. Akhmetshin et al., Phys. Lett. B 415 (1997) 445.
\bibitem{phietap2}
 R.R. Akhmetshin et al., Phys. Lett. B 473 (2000) 337.
\bibitem{phietap3}
 R.R. Akhmetshin et al., Phys. Lett. B 494 (2000) 26.
\bibitem{phietap4}
 V.M. Aulchenko et al., Pisma Zh. Eksp. Teor. Fiz. 69 (1999) 87.
\bibitem{phietap5}
 V.M. Aulchenko et al.,  Zh. Eksp. Teor. Fiz. 97 (2003) 28.
\bibitem{phiomp1} 
 M.N. Achasov et al., Phys. Lett. B 449 (1999) 122.
\bibitem{phiomp2} 
 M.N. Achasov et al., Nucl. Phys. B 569 (2000) 158.
\bibitem{phiomp3} 
 V.M. Aulchenko et al., Zh. Eksp. Teor. Fiz. 90 (2000) 1067.
\bibitem{phi4pic}
 R.R. Akhmetshin et al., Phys. Lett. B 491 (2000) 81.
\bibitem{rho4pic}
 R.R. Akhmetshin et al., Phys. Lett. B 475 (2000) 190.
\bibitem{snd-3pi4}
 M.N. Achasov et al., Phys. Rev. D 68 (2003) 052006.
\bibitem{kmd-pioee}
 R.R. Akhmetshin et al., Phys.Lett. B 503 (2001) 237. 
\bibitem{olya}
  L.M. Barkov et al., Nucl. Phys. B 256 (1985) 365.
\bibitem{snd-2pi}
 M.N. Achasov et al., Zh. Eksp. Teor. Fiz. 128 (2005) 1201,
 M.N. Achasov et al., these proceedings
\bibitem{tau04}
 S.I. Eidelman, in Proceedings of 8th International Workshop on $\tau$ 
 Lepton Physics (Tau 04), Nara, Japan, 14-17 Sep, 2004, Ed. by T. Ohshima,
 H. Hayashii, Nucl. Phys. (Proc. Suppl.) B 144 (2005) 223.
\bibitem{hep05}
 I.B. Logashenko et al., in Proceedings of International European Conference
 on High Energy Physics, Lisboa, Portugal, July 21-27, 2005
\bibitem{hadr05}
 A.L. Sibidanov et al., in Proceedings of 11 International Conference on 
 Hadron Spectroscopy, Rio-De-Janeiro, Brazil, August 21-26, 2005
\bibitem{kmd-2pi1}
 R.R. Akhmetshin et al, Phys. Lett. B 527 (2002) 161, \\
\bibitem{kmdrean}
 R.R. Akhmetshin et al, Phys. Lett. B 578 (2004) 285. 
\bibitem{kmd-2pi2}
 V.M. Aulchenko et al., Pisma Zh. Eksp. Teor. Fiz. 82 (2005) 841.
\bibitem{snd-3pi1}
 M.N. Achasov et al., Phys. Rev. D 63 (2001) 072002.
\bibitem{snd-3pi2}
 M.N. Achasov et al., Phys. Rev. D 65 (2002) 032002.
\bibitem{snd-3pi3}
 M.N. Achasov et al., Phys. Rev. D 66 (2002) 032001.
\bibitem{kmd-3pi2}
 R.R. Akhmetshin et al., Phys. Lett. B 434 (1998) 426.
\bibitem{kloe-3pi}
 A. Aloisio et al., Phys. Lett. B 561, 55 (2003); Erratum-ibid. B 609  (2005)
 449.
\bibitem{tomp3pi}
 N.N. Achasov, A.A. Kozhevnikov,  G.N. Shestakov, Phys. Lett. 50 B (1974) 448.\\
 N.N. Achasov, et al.,  Yad Fiz 23 (1976) 320.
\bibitem{pi3epif}
 D.A. Epifanov, in Proceedings of DA$\Phi$NE 2004: Workshop on Physics at
 Meson Factories, Rome, Frascati, Italy, 7-11 Jun., 2004,
 Frascati 2004, DA$\Phi$HNE 2004, p.389.
\bibitem{kmd-phi}
 R.R. Akhmetshin et al., Phys. Lett. B 364 (1995) 199.
\bibitem{kmd-3pi3}
 R.R. Akhmetshin et al., Phys. Lett. B 476 (2000) 33.
\bibitem{pi3gorb}
 D.A. Gorbachev, Proceedings of the 8th International Workshop on
 Meson Production, Properties and Interaction (MESON 2004), Krakow, Poland,
 4-8 June, 2004, Ed. by S. Kistryn, A. Magiera, H. Machner, C. Guaraldo, 
 Int. J. Mod. Phys. A 20 (2005) 584.
\bibitem{snd-pi0g2}
 M.N. Achasov et al., Eur. Phys. J. C 12, (2000) 25.
\bibitem{snd-etag3pc}
 M.N. Achasov et al., Zh. Eksp. Teor. Fiz. 117 (2000) 22.
\bibitem{snd-etag7g}
 M.N. Achasov et al., Pisma Zh. Eksp. Teor. Fiz. 72 (2000) 411.
\bibitem{kmd-etag3pc}
 R.R. Akhmetshin et al., Phys. Lett. B 460 (1999) 242.
\bibitem{kmd-kkc}
 P.A. Lukin,  in Proceedings of DA$\Phi$NE 2004: Workshop on Physics at
 Meson Factories, Rome, Frascati, Italy, 7-11 Jun., 2004,
 Frascati 2004, DA$\Phi$HNE 2004, p.381.
\bibitem{olya-kkc}
 P.M. Ivanov, et al., Phys. Lett. B 107 (1981) 297.
\bibitem{kmd-ksl}
 R.R. Akhmetshin et al., Phys. Lett. B 466 (1999) 385, Erratum-ibid. B 508
 (2001) 217.
\bibitem{kmd-ksl2}
 E.V. Anashkin et al.., Yad. Fiz. 65 (2002) 1255. 
\bibitem{kmd-ksl3}
 R.R. Akhmetshin et al., Phys. Lett. B 551 (2003).
\bibitem{kmd4p}
 R.R. Akhmetshin et al., Phys. Lett. B 466 (1999) 392.
\bibitem{cleo4p}
 K.W. Edwards et al., Phys. Rev. D 61 (2000) 072003.
\bibitem{snd-4pi}
 M.N. Achasov et al., Zh. Eksp. Teor. Fiz. 123 (2003) 899.
\bibitem{dm2-4pi}
 D. Bisello et al., Nucl. Phys. (Proc. Suppl.) B 21 (1991) 111.
\bibitem{kmd-4pi}
 R.R. Akhmetshin et al., Phys. Lett. B 595 (2004) 101.
\bibitem{snd-ppg}
 M.N. Achasov et al., Phys. Lett. B 486(2000) 29.
\bibitem{kmd-ppg}
 R.R. Akhmetshin et al., Phys. Lett. B 562(2003) 173.
\bibitem{cleo2}
 S. Anderson et al., Phys. Rev. D 61 (2000) 112002.
\bibitem{aleph2}
 S. Schael et al., Phys. Rept. 421 (2005) 191.
\bibitem{ospk72}
 D. Benaksas et al., Phys. Lett. B 39 (1972) 289.
\bibitem{spec95}
 R. Wurzinger et al., Phys. Rev. C 51 (1995) 443.
\bibitem{cbar1}
 C. Amsler et al., Phys. Lett. B 327 (1994) 425. 
\bibitem{cbar2}
 C. Amsler et al., Phys. Lett. B 311 (1993) 362.
\bibitem{aste}
 P. Weidenauer et al., Z. Phys C 59 (1993) 387.
\bibitem{olyaomega}
 L.M. Kurdadze et al., Pisma Zh. Eksp. Teor. Fiz. 36 (1982) 221.
\bibitem{cntr76}
 J. Keyne et al., Phys. Rev. D 14 (1976) 28.
\bibitem{dm1}
 A. Cordier et al., Nucl. Phys. B 172 (1980) 13.
\bibitem{nd}
 S.I. Dolinsky et al., Phys. Rep. 202 (1991) 99.
\bibitem{cmdgom}
 L.M. Barkov et al, Pisma Zh. Eksp. Teor. Fiz. 46 (1987) 132.
\bibitem{dm1-2}
 A. Quenzer et al., Phys. Lett. B 76 (1978) 512.
\bibitem{aspk72}
 B.N. Ratcliff et al., Phys. Lett. B 38 (1972) 345.
\bibitem{aspk71}
 H.J. Behrend et al., Phys. Rev. Lett. 27 (1971) 61.
\bibitem{omeg98}
 D. Barberis et al., Phys. Lett. B 432 (1998) 466.
\bibitem{mpsf86}
 T.F. Davenport et al., Phys. Rev. D 33 (1986) 2519.
\bibitem{arg85}
 H. Albercht et al., Phys. Lett. B 153 (1985) 343.
\bibitem{aems82}
 M.W. Arenton et al., Phys. Rev. D 25 (1982) 2241.
\bibitem{olya78}
 A.D. Bukin et al., Yad. Fiz. 27 (1978) 976.
\bibitem{cntr74}
 H.J. Besch et al., Nucl. Phys. B 70 (1974) 257.
\bibitem{ospk74}
 G. Cosme et. al., Phys. Lett. B 63 (1976) 352.
\bibitem{ospk71}
 E.V. Balakin et. al., Phys. Lett. B 34 (1974) 328.
\bibitem{ospk70}
 J.C. Bizot et. al., Phys. Lett. B 32 (1970) 416.
\bibitem{hbc76}
 G.R. Kalbfleish, R.C. Strand, J.W. Chapman, Phys. Rev. D 13 (1976) 22.
\bibitem{hbc74}
 A.J. de Groot, Nucl. Phys. B 74 (1974) 77.
\bibitem{hbc66}
 J.S. Lindsey, G. Smith, Phys. Rev. D 147 (1966) 913.
\bibitem{hbc78}
 M.J. Losty, Nucl. Phys. B 133 (1978) 38.
\bibitem{hbc771}
 H.Laven et al., Nucl. Phys. B 127 (1977) 43.
\bibitem{hbc772}
 L. Lyons, A.M. Cooper, A.G. Clark, Nucl. Phys. B 125 (1977) 207.
\bibitem{hbc72}
 M. Aguilar-Benitez et al., Phys. Rev. D 6 (1972) 29.
\bibitem{ospk76}
 G. Parrour et al., Phys. Lett. B 63 (1976) 357.
\bibitem{olya83}
 L.M. Kurdadze et al., Zh. Eksp. Teor. Fiz. 38 (1983) 306.
\bibitem{cntr77}
 D.E. Andrews et al., Phys. Rev. Lett. 38 (1977) 198.
\bibitem{snd-2mu}
 M.N. Achasov et al., Phys. Rev. Lett. 86 (2001) 1698.
\bibitem{kloe-ll}
 F. Ambrosino et al., Phys. Lett. B 608 (2005) 199.
\end{thebibliography}
\end{document}